\documentclass[sn-nature]{sn-jnl}

\usepackage{graphicx}%
\usepackage{multirow}%
\usepackage{amsmath,amssymb,amsfonts}%
\usepackage{amsthm}%
\usepackage{mathrsfs}%
\usepackage[title]{appendix}%
\usepackage{xcolor}%
\usepackage{textcomp}%
\usepackage{manyfoot}%
\usepackage{booktabs}%
\usepackage{algorithm}%
\usepackage{algorithmicx}%
\usepackage{algpseudocode}%
\usepackage{listings}%

\begin{document}

\title[Article Title]{Immiscible to miscible quenching instabilities in two-dimensional 
binary Bose-Einstein condensates}


\author*[1]{\fnm{Lauro} \sur{Tomio}}\email{lauro.tomio@unesp.br}

\author[1]{\fnm{S.} \sur{Sabari}}\email{ssabari01@gmail.com}

\author[2]{\fnm{A.} \sur{Gammal}}\email{gammal@if.usp.br}

\author[1]{\fnm{R. K. } \sur{Kumar}}\email{kishor.bec@gmail.com}

\affil*[1]
{\orgdiv{Instituto de F\'{i}sica Te\'orica}, \orgname{Universidade Estadual Paulista (UNESP)}, \\
\postcode{01140-070} \city{S\~ao Paulo}, \state{SP}, \country{Brazil}}

\affil[2]
{Instituto de F\'{i}sica, Universidade de S\~{a}o Paulo (USP),  05508-090 S\~{a}o Paulo, Brazil}

\abstract{\small Immiscible to miscible quenching transitions (IMQT) in homogeneous Bose-Einstein 
condensate are investigated, considering rubidium isotopes $^{85}$Rb and $^{87}$Rbconfined in a 
two-dimensional  (2D) circular box, under two different initial configurations. These IMQT instabilities, 
triggered by sudden reductions in the two-body interspecies scattering length $a_{12}$, are explored 
under two distinct initialconditions, highlighting the critical role of nonlinear dynamics in their evolution. 
The numerical simulations indicate that the instability dynamics are primarily driven by the production 
of large vortices and the propagation of sound waves (phonons), with sound wave excitations 
prevailing in the long-term evolution. The compressible and incompressible parts of the kinetic 
energy spectra, in terms of the wave number $k$, are confronted with the classical Kolmogorov 
scaling, $k^{-5/3}$ for turbulence, which is observed in the onset of instabilities. Before reaching 
the ultraviolet dissipation region at small scales, the IMQT spectra exhibit a bottleneck effect, indicating 
a clear departure from classical scaling behavior. In the time asymptotic miscible regime, it is observed 
that the vorticity and sound-wave production remain practically stable. In this regime, for both cases 
investigated, a linear relation is also recognized between the miscibility parameter and the initial IMQT 
configuration.
}



\maketitle

\section{Introduction}\label{sec1}
Following a recent study on binary instability, we try to distinguish the cases in which
external linear forces are responsible for the dynamics, from the cases in which the dynamical 
instabilities are due to sudden variations of the nonlinear interactions~\cite{2025Kumar}. 
As detailed in this reference, by considering numerical simulations for a binary mixture of
rubidium isotopes, $^{85}$Rb and $^{87}$Rb, the main motivation was to provide numerical
simulations leading to different kinds of quantum instabilities, with turbulence and vortex generation,
that could be generated by dynamical solutions of the nonlinear Gross-Pitaevskii (GP) formalism. 
Within this study, one of the objectives was also to verify the possible emergence 
of scaling laws for the compressible and incompressible parts of the kinetic energy spectra, in terms
of the wave-number $k$, to be compared with the classical Kolmogorov scalings, $k^{-5/3}$ and 
$k^{-3}$, for turbulence~\cite{Kolmogorov1941,1995Frisch,1986Donnelly}. 
In the third approach discussed in Ref.~\cite{2025Kumar}, we investigate the dynamical response 
of the system under an immiscible to a miscible quenching transition (IMQT), 
implemented by a sudden reduction in the ratio between the inter- and intraspecies 
scattering lengths near the transition threshold. For that, the dynamics was 
explored by considering two different initial conditions for the space configuration 
and quenching transition. The particular interest in this case follows previous studies on phase 
separation and modulation instabilities with two-component atomic 
systems~\cite{2024Bayocboc,Mukherjee2020,2016Eto,2004Kasamatsu,2019Thiruvalluvar}.
As demonstrated in Ref.~\cite{2025Kumar}, the IMQT induces numerous vortex dipoles
and turbulent flow in the condensates, in addition to several interesting differences
from the cases where external linear forces are assumed.
To investigate in more detail the dynamics, we have analyzed the corresponding 
compressible and incompressible kinetic energy parts of the spectrum, following an 
approach provided in~\cite{2022Bradley}. 
Based on classical fluid mixture studies, the compressible component corresponds to density
fluctuations and the emission of sound waves (phonons), while the incompressible component is
linked to vortex dynamics and rotational motion. Analyzing these components can reveal possible 
universal scaling laws, potentially establishing consistency with the classical 
Kolmogorov scaling for turbulence, as discussed in Ref.~\cite{2013Reeves}.
The emergence of turbulence in Bose–Einstein condensates (BECs), 
known as quantum turbulence (QT), was first reported in Ref.~\cite{Bagnato2009}.
An updated bibliography on QT, including its experimental status and recent theoretical 
analyses, can be found in Refs.~\cite{2016Tsatsos,2025Moreno,2024Madeira}, as well as in 
recent studies by some of us on vortex generation in binary BECs~\cite{2023Silva,Lauro2024,Sabari2024}. 

In the present study, we aim to extend our previous analysis of the IMQT by considering 
different initial conditions from those explored in Ref.~\cite{2025Kumar}, while employing 
the same numerical approach.
To this end, we use similar analytical techniques, beginning with an examination of the spectrum 
for the corresponding Kolmogorov scaling. For this purpose, the behavior of the compressible and 
incompressible parts of the fluid is examined across scales—from large to small—during time 
intervals when instabilities may emerge, allowing a comparison with the classical counterpart.
At later times, however, the Kolmogorov-like spectral analysis becomes less useful for such a 
comparison due to the absence of viscosity and the presence of quantized vortices. In this 
regime, we instead investigate a residual effect of the initial conditions on the dynamics, 
which was previously noted in Ref.~\cite{2025Kumar}.

\section{Mean-field model for binary BEC in a uniform circular box}\label{sec2}
For the coupled BEC system, we assume the atomic rubidium isotopes $^{85}$Rb and $^{87}$Rb,
respectively identified as species $i=$1 and 2, with masses $m_i$. Both species are assumed 
having the same number of atoms $N_i\equiv N$, 
initially confined in strong pancake-shaped harmonic traps with fixed aspect ratios 
$\lambda\equiv \lambda_i=\omega_{i,z}/\omega_{i,\perp}$, where $\omega_{i,z}$ and 
$\omega_{i,\perp}$ are, respectively, the longitudinal and transverse trap frequencies
of the species $i$. 
For convenience and computational purposes, the GP 2D formalism is cast in a 
dimensionless format, using the original harmonic trap parameters, with energy, time, 
and length units given, respectively, by $\hbar \omega_{\perp}$, $\omega_{\perp}^{-1}$, 
and $l_{\perp}\equiv \sqrt{\hbar/(m_1\omega_{\perp})}$, 
with the first species being used as the reference for the length unit. 
Correspondingly, the space and time variables are scaled as   
${\bf r}\to l_\perp {\bf r}$ and $t\to t/\omega_{\perp}$, when converting to dimensionless units.
In applying this dimensionless 3D-to-2D reduction to a mass-imbalanced binary system, we follow an 
analogous procedure to that described in Ref.~\cite{2019Kumar}.
The coupled 2D GP equation, for the wave-function components 
$\psi_i\equiv\psi_i(x,y;t)$, 
normalized to one, $\int_{-\infty}^{\infty}dx\,dy\,|\psi _{i}|^{2}=1$, is given by 
\begin{eqnarray}
\mathrm{i}\frac{\partial \psi_{i}  }{\partial t }
&=&{\bigg\{}\frac{-m_{1}}{2m_{i}}\left(\frac{\partial^2}{\partial x^2} + 
\frac{\partial^2}{\partial y^2}\right)
+V_i(x,y) +\sum_{j=1,2}g_{ij}|\psi_{j} |^{2}{\bigg\}}
\psi_{i}  
\label{2d-2c}, 
\end{eqnarray}
where the 2D nonlinear strengths $g_{ij}$ refer to the two-body interactions, 
related to the $s-$wave two-body 
scattering lengths $a_{ij}=a_{ji}$, given by 
\begin{eqnarray}
g_{ij}\equiv \sqrt{2\pi\lambda}
\frac{m_1 a_{ij} N_j}{m_{ij}l_\perp},
\label{par}
\end{eqnarray}
where $\mu_{ij} \equiv m_im_j/(m_i+m_j)$ is the reduced mass, and 
we assume $a_{11}=a_{22}$, with the same number of atoms for both species, $N_{1}=N_{2}$.
The initially assumed 2D harmonic trap potential, identical for both species, is modified by a  
uniform circular box with fixed radius $R$ and height $V_0$, 
much larger than the chemical potential $\mu$. It is given by
\begin{eqnarray}
V_{i}(x,y) 
&=&\left\{ \begin{array}{l}
V_0,\;{\rm for}\;\;\sqrt{x^2+y^2}>R,\\
0,\;\;{\rm for}\;\;\sqrt{x^2+y^2}\le R.
\end{array}\right.
\label{2Dtrap}
\end{eqnarray}
Next, in the present approach, the length unit will be adjusted to  $l_\perp = 1\mu$m$ \approx 
1.89\times 10^4 a_0$, where $a_0$ is the Bohr radius, with $a_{ij}$ given conveniently in terms of $a_0$.
Both components are assumed to have an equal number of atoms, $N=3.5\times10^4$. and equal intraspecies
interactions, $a_{11}=a_{22}$. Therefore, relying on the actual possibilities to alter
the two-body scattering lengths~\cite{1999Timmermans}, the interspecies $a_{12}$ will be
a key parameter for controlling the miscibility of the coupled system, as will be shown.

The immiscible regime condition corresponds to $g_{12}^2 > g_{11} g_{22}$ in~\eqref{par}. 
Therefore, for $a_{ii}>0$, the threshold parameter $\delta$ is given by
\begin{equation}
\delta\equiv \frac{a_{12}}{a_{11}} > \frac{2\sqrt{m_1m_2}}{m_1+m_2}.
\label{delta}\end{equation}
The right side of the above relation is the mass-dependent critical value for the immiscible-to-miscible 
transition of the homogeneous mixture. In the present case, considering the binary mixture of $^{85}$Rb and $^{87}$Rb,
this critical value deviates from one only in the fifth decimal digit, and one can assume the same relation as
the equal-mass one, with $\delta=1$ being the transition point from immiscible to miscible regime.
The corresponding IMQT refers to the sudden reduction of $a_{12}$, from $a_{12}^{(0)}$ 
(which is assumed to be larger than the intraspecies $a_{ii}$, at $t=0$) to a final $a_{12}$ 
(smaller than  $a_{ii}$, at any $t>0$), 
with $a_{11}=a_{22}$ kept fixed along the dynamics. This IMQT is defined  by
\begin{equation}
\Delta\delta\equiv \frac{(a_{12}^{(0)}-a_{12})}{a_{11}} \equiv \frac{\Delta (a_{12})}{a_{11}}.
\label{IMQT}\end{equation}

To improve the analysis for the immiscible to miscible transition, performed in our previous study~\cite{2025Kumar}, 
in the present investigation we consider a larger transition from immiscible to miscible regimes. In this case,
$a_{ii}=90a_0$ was chosen for the intraspecies, with $a_{12}=100a_0$ and $a_{12}=110a_0$, for two
independent initial immiscible configurations. Before starting the evolution, at $t=0$, the first has the 
coupled species sharing three non-overlapping regions; with the second, sharing symmetrically two 
non-overlapping regions, as will be clarified.  For the final miscible configuration in the dynamics, 
in both cases we assume $a_{12}=60a_0$. 
Therefore, in the first case, called {\it central} considering that we assume one of the species
in the central intermediate region, with the initial $a_{12}=100a_0$ quenched to $60a_0$, the IMQT 
is given by $\Delta\delta = 4/9$, whereas in the second case, called {\it axial}, with the 
initial $a_{12}=110a_0$ suddenly changed to $a_{12}=60a_0$, the IMQT is $\Delta\delta =5/9$.
Respectively, in Ref.~\cite{2025Kumar}, an IMQT given by $\Delta\delta = 0.27$ was assumed for the  
central case; and $\Delta\delta =0.30$, for the axial case.

An improved definition for the miscibility, to be considered in the dynamics analyses, is obtained by  
using the overlap between the densities, as the ones defined in Refs.~\cite{2017kumar} and 
\cite{Mukherjee2020}. For the present 2D case, we assume the prescription presented in~\cite{Mukherjee2020}
for the overlap between the densities, which is given by
\begin{eqnarray}
\Lambda = \frac{\left[\int \vert \psi_1 \vert^2 \vert \psi_2 \vert^2 \ dx dy \right]^2}{ \left(
\int {\vert \psi_1 \vert^2} \ dxdy \right) \left( \int { \vert \psi_2 \vert^2} \ dxdy \right) } 
.\label{eta_new}\end{eqnarray}
So, the complete overlap between the two densities implies $\Lambda=$1, and  $\Lambda=$0 in the other extreme
of full-immiscible case. 

For the numerical approach to solve the corresponding GP equation~(\ref{2d-2c}), we have used the split-step 
Fourier method. The numerical simulations are performed using an 800$\times$800 square grid with box 
length $L\equiv L_x=L_y=48$ and time step $\Delta t=10^{-3}$. 
For the spectra numerical analysis, the corresponding wavenumber infrared limit is $k_L=2\pi/L$.
The given space and time units are, respectively, $l_\perp=1\mu m$ and $\omega_\perp^{-1}$. 
The healing length is given by $\xi = \hbar/\sqrt{m\mu} \sim $0.1$ \mu$m, which is considered the same
for both species.
Since we consider the same number of atoms and the length of scattering, only the small mass-difference 
($m_1$/$m_2$=0.98) plays a role in determining the chemical potential. 
We should be aware that in the present simulations, for convenience in computing time reduction, we have 
assumed the number of atoms ($N=3.5\times10^4$) smaller than in Ref.~\cite{2025Kumar} ($N=2\times10^6$), 
implying in different box sizes. However, as it will be evidenced, these new assumptions have not affected 
qualitatively the results, which are mainly due to the new IMQT interaction parameters.

In order to study the energy spectrum and corresponding similarity of the results with the classical Kolmogorov 
scalings for turbulence, below we present the main necessary equations. 
For more details, see Ref.~\cite{2025Kumar}. 
With the 2D variables defined as ${\pmb \rho}\equiv (x,y)$, the
time-dependent density is given by $n_i({\pmb \rho},t)\equiv|\psi_i({\pmb \rho};t)|^2$, having 
the density-weighted velocity field defined by 
${\bf u}_i({\pmb \rho},t)\, \equiv\,\sqrt{n_i({\pmb \rho},t)} {\bf v}_i({\pmb \rho},t) $.
For our following spectral analysis, the kinetic energy is decomposed into compressible and incompressible parts. 
For that, the density-weighted velocity field is given by two terms, 
${\bf u}_i({\pmb \rho},t)\, \equiv\, {\bf u}_i^I({\pmb \rho},t) + {\bf u}^C_i({\pmb \rho},t)$, where ${\bf u}_i^I$ 
is the incompressible ($I$) field that satisfies $\nabla .{\bf u}_i^I = 0$, with the compressible ($C$) one, 
${\bf u}_i^C$, satisfying $\nabla \times{\bf u}^C_i = 0$. Following that, we have the kinetic energy decomposed
as $E_i^K=E_i^I+E_i^C$; respectively, associated to incompressible and compressible kinetic energy parts,
given by
\begin{align}
E_i^{C,I} = \frac{m_1}{2m_i} \int d^2{\pmb \rho} |{\bf u}^{C,I}_i({\pmb \rho},t)|^2
\end{align}
By calculated the time evolution of 
both compressible and incompressible  parts of kinetic energies, we are able to investigate 
the corresponding contribution of sound wave production (compressible) and vorticity (incompressible).
In momentum space, with ${\bf k}$ being the wave vector, 
these  kinetic energy terms can be expressed by
\begin{align}
E_i^{C,I} =\frac{m_1}{2m_i}  \int d^2{\bf k} 
\left|\frac{1}{2\pi} \int d^2{\pmb \rho}\, 
 e^{-i{\bf k}.{\pmb \rho}} {\bf u}_i^{C,I}({\pmb \rho})\right|^2.
\end{align}
Within this procedure, we have the spectral density in 2D $k-$space in the integrand of the above expression,
which is integrated in polar coordinates $(k,\phi_k)$ (where $k=\sqrt{k_x^2+k_y^2}$).  The final expression for
the compressible and incompressible kinetic energy spectra, in $k$ space, are obtained by integrating over 
the azimuthal angle, as
\begin{align}
{\cal E}_i^{C,I}( k) = \frac{m_1\; k}{2m_i} \int_0^{2\pi} d\phi_k \left|
\frac{1}{2\pi} \int d^2{\pmb \rho}\, 
 e^{-i{\bf k}.{\pmb \rho}} {\bf u}^{C,I}_i({\pmb \rho})
\right|^2. 
\end{align}

\section{Main results}\label{sec3}
In our numerical approach, the initial ground state of the system is obtained by propagating Eq.~(\ref{2d-2c}) 
in imaginary time, considering the interspecies interaction $a_{12}$ larger than the intraspecies  
$a_{11}=a_{22}$, which is followed by dynamical simulations in real time, with a sudden reduction of $a_{12}$ to $a_{12}<a_{ii}$.
To study the dynamics, we consider two kinds of initial configurations. The first, by having the homogeneous 
initial configuration with the two species separated into three non-overlapping spatial domains, inside the circle.
In such a case, initially in an immiscible configuration, the coupled system shares two interface
regions before starting the time evolution going from an immiscible to a miscible regime. 
In Ref.~\cite{2025Kumar} this was named a {\it tennis-ball-shaped} configuration (or tripartite spatial configuration)
In the other configuration, called an{\it axial-shaped} configuration, we assume that the two species 
are axially separated, such that only one interface exists to initially separate the components. 
These two cases follow the ones previously assumed in Ref.~\cite{2025Kumar}, but with different sudden 
variations in the interspecies interactions, such that we can verify the corresponding effect in the 
immiscible to miscible transition.
\begin{figure*}[tb]
\begin{center}
\includegraphics[width=13.5cm,height=4cm]{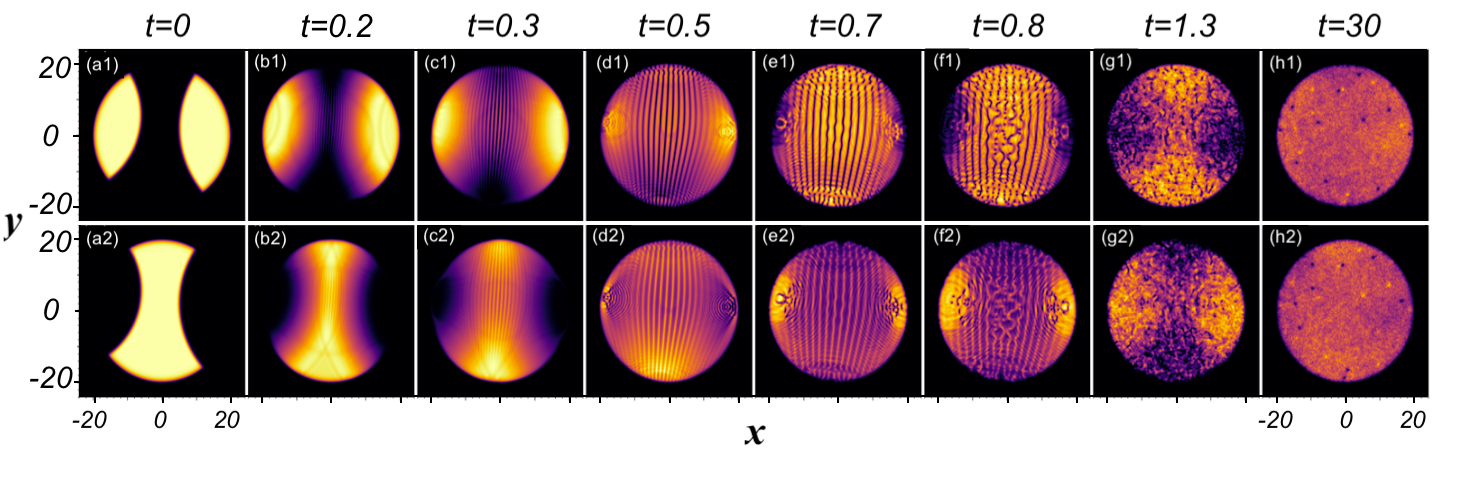}
\caption{For different time snap shots indicated at the top, the upper panels (a$_1$)-(h$_1$) 
show density profiles of 
the first component ($^{85}$Rb), with the lower panels (a$_2$)-(h$_2$) showing the density 
profiles of the second 
component ($^{87}$Rb). In the initial configuration, at  $t=0$ (a$_1$), we have $a_{12}=100a_0$, 
which is suddenly
 quenched to $a_{12}=60a_0$, before starting the evolution as shown from $t=0.2$ (b$_1$). 
 As $a_{11}$ and $a_{22}$ are fixed at $90a_0$, the corresponding IMQT is $\Delta\delta=4/9$.
 The space and time units are, respectively, $l_\perp$ and $1/\omega_\perp$.
 }
\label{fig1}
\end{center}
\end{figure*}

\begin{figure}[tb]
\begin{center}
\includegraphics[width=0.6\textwidth]{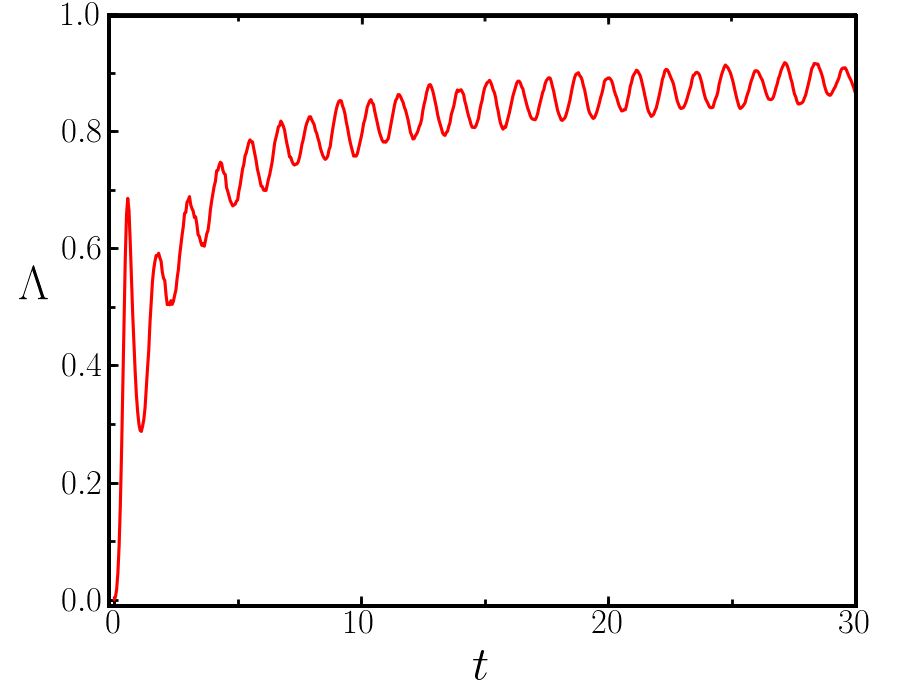}
\caption{Time evolution of the density overlap parameter $\Lambda$, corresponding to 
Fig.~\ref{fig1}, showing how the binary system evolve from an immiscible to a miscible regime. 
For the IMQT, with $a_{11}=a_{22}=90a_0$, the interspecies interaction is suddenly changed 
from $a_{12}=100a_0$ to $a_{12}=60a_0$, before starting the evolution. 
$\Delta\delta=4/9$ is the corresponding IMQT, in this case. 
In the asymptotic miscible regime, one can verify that the oscillating frequency is 
$\nu_\Lambda\approx 0.82\omega_\perp$. The time unit is $\omega_\perp^{-1}$.}
\label{fig2}
\end{center}
\end{figure}

\subsection{Three spatial domains trap configuration}
Initially, the two species are prepared in three spatial domains, as shown by the
densities in panels (a$_1$) and (a$_2$) of Fig.~\ref{fig1}, with $^{87}$Rb, located in
the central part. To obtain a phase-separated binary mixture, we consider $a_{12}=100a_0$ 
slightly  larger than the intraspecies interactions, $a_{11}=a_{22}=90a_0$. 
The dynamics is followed by evolving the ground state in real-time, with  the 
interspecies scattering length interaction quenched to $a_{12}=60a_0$ to produce phase mixing.
Therefore, in the present study, we assume the initial immiscible configuration \eqref{delta}
given by $\delta=10/9$, suddenly changed to $\delta=6/9$ when starting the time
evolution. It implies that the related IMQT \eqref{IMQT} is $\Delta\delta =4/9$, 
much larger than the transition $\Delta\delta = 0.27$ assumed in Ref.~\cite{2025Kumar} 
for the same kind of initial spatial configuration.
 
By comparing Fig.~\ref{fig1} with figure~8 of Ref.~\cite{2025Kumar}, we observe that 
the results in the interference fringes are similar along the time evolution. 
However, all the dynamics start at a reduced time in the present case, consistent 
with the actual stronger IMQT, which is about 1.65 larger than the one considered 
in Ref.~\cite{2025Kumar}. Besides the fact that we are now assuming species 2 
($^{87}$Rb) located at the center part, differently from Ref.~\cite{2025Kumar}, we 
attribute such differences mainly to the larger IMQT that we are assuming here, 
in view of the small mass difference between the species. 
In the asymptotic limit ($t>20$), similar behaviors can be visually verified from the 
densities, which indicate an approximately constant number of vortices for $t>20$. 
The memory of the initial configuration remains basically in the oscillating frequency
of the coupled densities, which can be observed in the time evolution of the IMQT 
parameter $\Lambda$ provided in Fig.~\ref{fig2}. This figure shows how the miscibility 
evolves in time when $a_{12}$ suffers a sudden reduction, with the system changing 
from an immiscible to a miscicle configuration, such that $\Delta\delta= 4/9$. 
With the assumption that with $t>20$ we have the asymptotic limit, we observe that 
$\Delta\delta=4/9$ implies in a constant density oscillation frequency 
$\nu_\Lambda\approx 5/6$ (units of $\omega_\perp$), whereas for $\Delta\delta= 0.27$ 
(assumed in Ref.~\cite{2025Kumar}), we have $\nu_\Lambda\approx 1/3$ (See figure~14 of
Ref.~\cite{2025Kumar}). The difference between the results is clearly related to the 
different initial IMQT assumptions in both cases. As verified by a direct comparison 
of both results, a larger initial IMQT implies in stronger constant oscillations, 
asymptotically, evidenced by the densities in the miscible regime of the coupled mixture. 
By considering both cases, we can roughly associate the frequency $\nu_\Lambda$ with 
the initial IMQT, by a linear relation, which in this case is given by
$\Delta\delta\approx 0.36\nu_\Lambda +0.15$. 
The factor 0.36 originates from the initial immiscible configuration (common to both 
cases, featuring two density interfaces). In contrast, the constant 0.15 corresponds 
to the threshold IMQT, which emerges in the zero-frequency limit 
(where $\nu_\Lambda \approx 0$ for $\Delta\delta \approx 0.15$).

\begin{figure*}[tb]
\begin{center}
\includegraphics[width=1.0\textwidth]{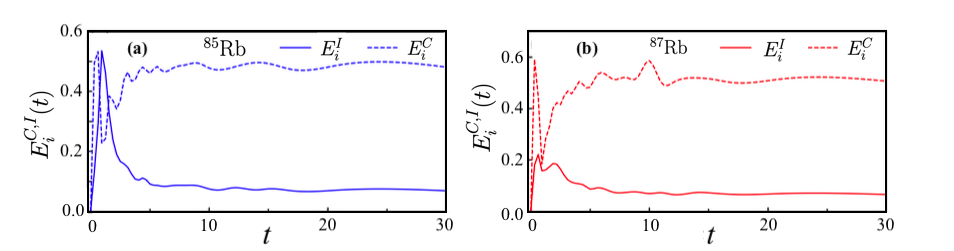}
\caption{Time evolution of the incompressible and compressible energies, 
corresponding to the dynamics presented in Fig.~\ref{fig1}, for the first (a) and second (b) components, 
during the immiscible to miscible transition.
The energy and time units are, respectively, $\hbar\omega_\perp$ and $1/\omega_\perp$.
}
\label{fig3}
\end{center}
\end{figure*}

The IMQT introduced in this study exhibits a faster response than that reported in Ref.~\cite{2025Kumar}. This difference is also evident in the time evolution of the 
compressible and incompressible energy components for both species, shown in 
Fig.~\ref{fig3}, compared to the upper frame of figure 9 of Ref.~\cite{2025Kumar} 
(where data for both species are combined in a single panel).
The number of vortices $N_v$ is quite large along the dynamics, as already being 
reported in Ref.~\cite{2025Kumar} for a similar case, remaining a non-zero constant
(of the order of 20) in the asymptotic limit. In our present study, $N_v$ is 
not explicitly shown, as it can be directly related with the behavior of the 
incompressible parts of the kinetic energies presented in Fig.~\ref{fig3}. 
In the asymptotic limit, one can also visually observe the remaining vortices by 
considering the panels (h$_i$) of Fig.\ref{fig1}, where $t>30$.
Notice that such asymptotic limit behavior starts before 
$t\approx 10$, whereas in Ref.~\cite{2025Kumar} it starts near $t\approx 20$. 
Fig.\ref{fig3} presents the two species separately: $^{85}$Rb in panel (a) and 
$^{87}$Rb in panel (b). For both species, the asymptotic compressible energy 
(linked to sound-wave production) dominates the incompressible energy by roughly 
a factor of five. Their temporal evolution diverges primarily near the 
onset or the instabilities ($t<5$), reflecting the influence of their different initial spatial 
configurations.

\begin{figure*}[tb]
\begin{center}
\includegraphics[width=1.0\textwidth]{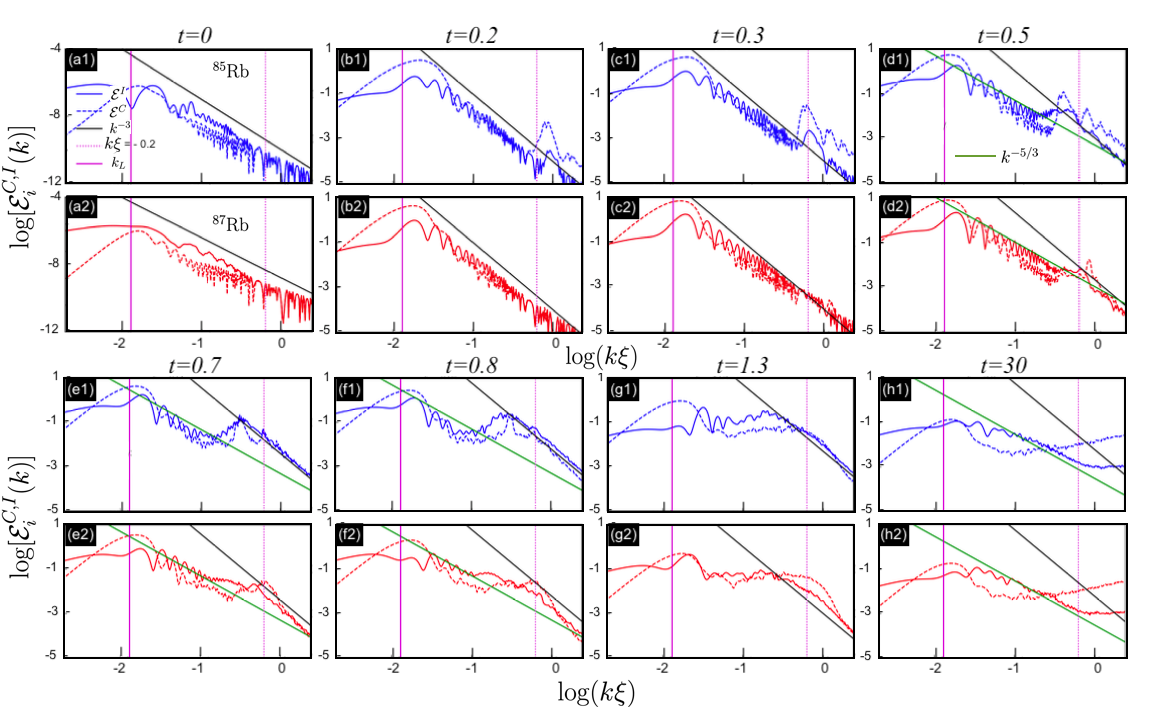}
\caption{Corresponding to the profiles shown in Fig.~\ref{fig1},
the incompressible and compressible kinetic energy spectra of $^{85}$Rb (species 1)
and $^{8t}$Rb (species 2) are presented in log-scale as functions of $\log(k\xi)$
(where $\xi$ is the healing length), for given time instants
(indicated at the top of the panels). With the conventions indicated in the
panel (a$_1$), the expected  $k^{-5/3}$ Kolmogorov behavior stated in (d1) is 
represented by the green lines, being more evident for $0.5\le t \le 0.8$. 
The vertical solid line refers to the infrared box limit ($k_L\approx 0.13$),
with the vertical dotted line indicating the start of the ultra-violet
branch region (obtained up to $k_{max}\xi\approx\pi/2$).
The length, energy and time units are, respectively, 
$l_\perp$, $\hbar\omega_\perp$ and $1/\omega_\perp$.
}
\label{fig4}
\end{center}
\end{figure*}

Finally, for this immiscible configuration — where the coupled system occupies three 
distinct spatial domains — we also observe that the Kolmogorov behavior discussed 
in Ref.~\cite{2025Kumar} emerges at earlier times, a consequence of the larger IMQT.
This can be observed by inspecting Fig.~\ref{fig4} [See, for example, panels (d$_i$)
and (e$_{i}$)] together with figure~10 of Ref.~\cite{2025Kumar}. 
A related feature is the bottleneck effect noted in Ref.~\cite{2007Lvov} and also 
discussed in~\cite{2025Kumar}. This effect is visible in panels (d$_i$) to (f$_i$) 
and is consistently more pronounced in the spectrum of the component located outside 
the center. In our case, this is $^{85}$Rb, whereas in Ref.~\cite{2025Kumar}, it is
$^{87}$Rb.

\subsection{Axial-shaped initial spatial configuration }
\begin{figure*}[tb]
\begin{center}
\includegraphics[width=13.5cm,height=4cm]{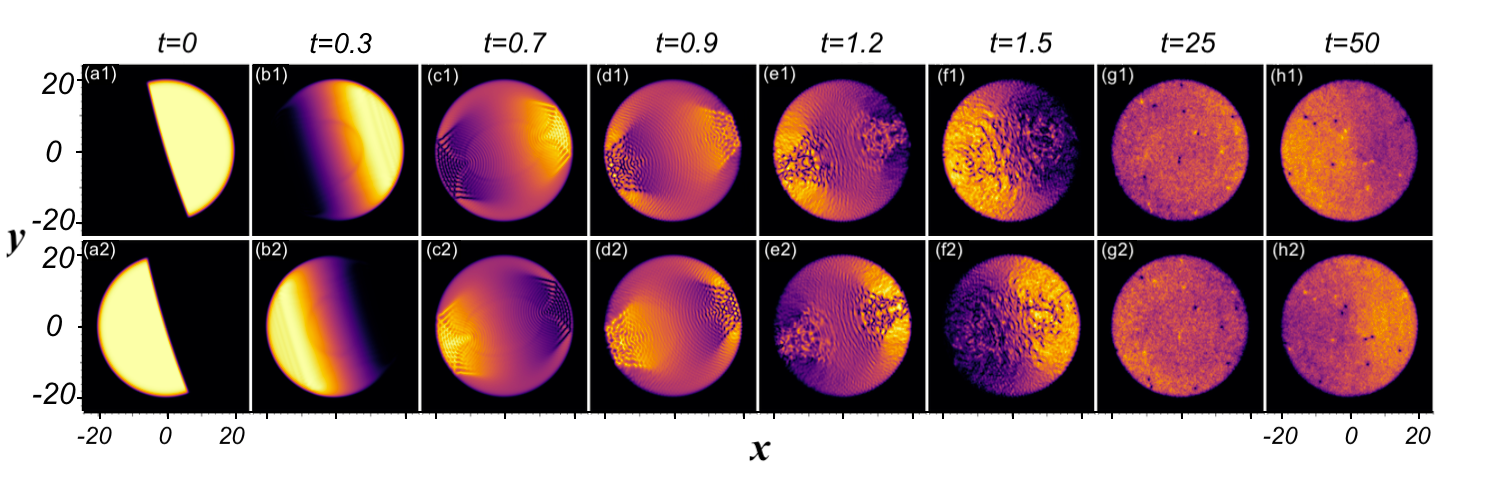}\vspace{-.5cm}
\caption{For different time snap shots indicated at the top, the upper panels (a$_1$)-(h$_1$) 
show density profiles of 
the first component ($^{85}$Rb), with the lower panels (a$_2$)-(h$_2$) showing the density profiles of the second 
component ($^{87}$Rb). In the initial configuration, at  $t=0$ (a$_1$), we have $a_{12}=110a_0$, which is suddenly
 quenched to $a_{12}=60a_0$ (kept fixed along the dynamics). As $a_{11}$ and $a_{22}$ are fixed at $90a_0$, 
 the corresponding IMQT is $\Delta\delta=5/9$.  The space and time units are, respectively, $l_\perp$ and 
 $1/\omega_\perp$.
 }
\label{fig5}
\end{center}
\end{figure*}

\begin{figure}[tb]
\begin{center}
\includegraphics[width=0.6\textwidth]{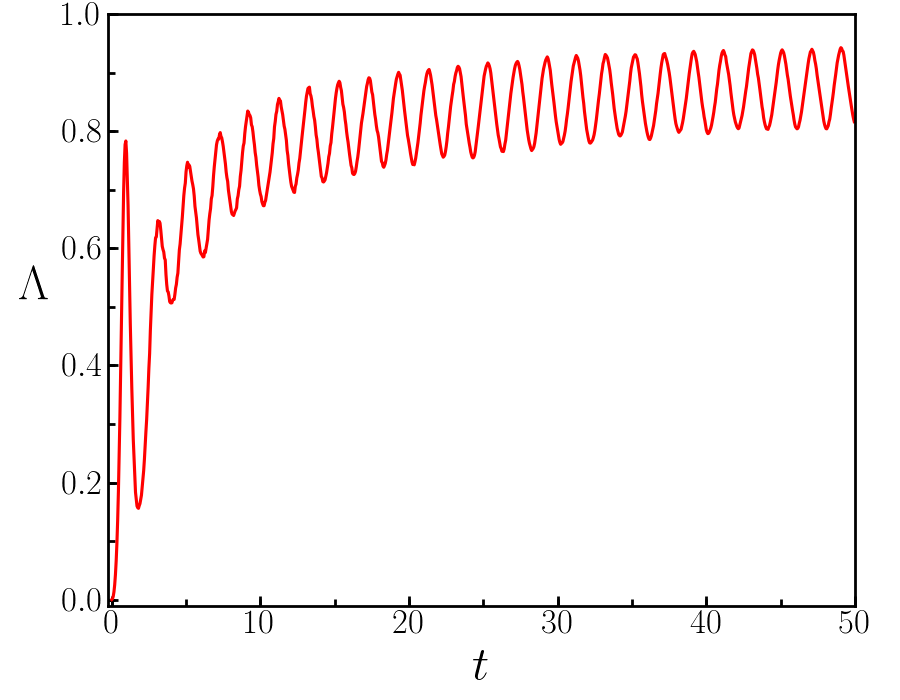}
\caption{Time evolution of the density overlap parameter $\Lambda$, corresponding to 
Fig.~\ref{fig5}, showing how the binary system evolve from an immiscible to a miscible 
regime. For the IMQT, with fixed $a_{11}=a_{22}=90a_0$, the interspecies interaction 
is suddenly changed from $a_{12}=110a_0$ to $a_{12}=60a_0$, which is kept along the 
time evolution. $\Delta\delta=5/9$ is the corresponding IMQT, in this case.
In the asymptotic miscible regime, one can verify that the oscillating frequency is 
$\nu_\Lambda\approx 0.5\omega_\perp$. The time unit is $\omega_\perp^{-1}$.}
\label{fig6}
\end{center}
\end{figure}

\begin{figure*}[tb]
\begin{center}
\includegraphics[width=1.0\textwidth]{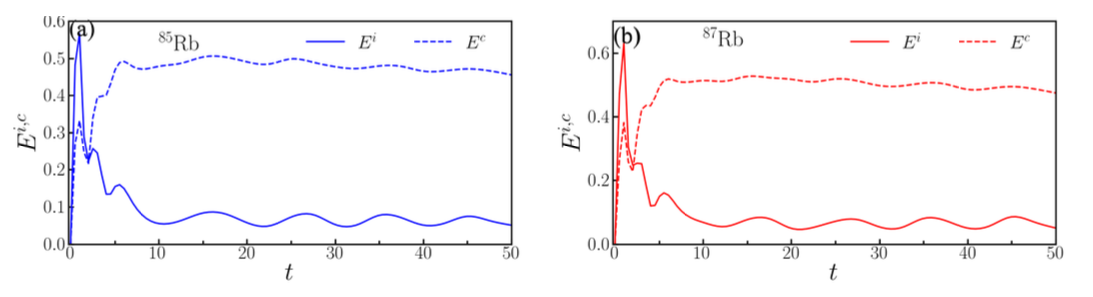}
\caption{
Time evolution of the incompressible and compressible energies, 
corresponding to the dynamics presented in Fig.~\ref{fig5}, for the first (a) and second (b) components, 
during the immiscible to miscible transition.
The energy and time units are, respectively, $\hbar\omega_\perp$ and $1/\omega_\perp$.}
\label{fig7}
\end{center}
\end{figure*}

\begin{figure*}[tb]
\begin{center}
\includegraphics[width=1.0\textwidth]{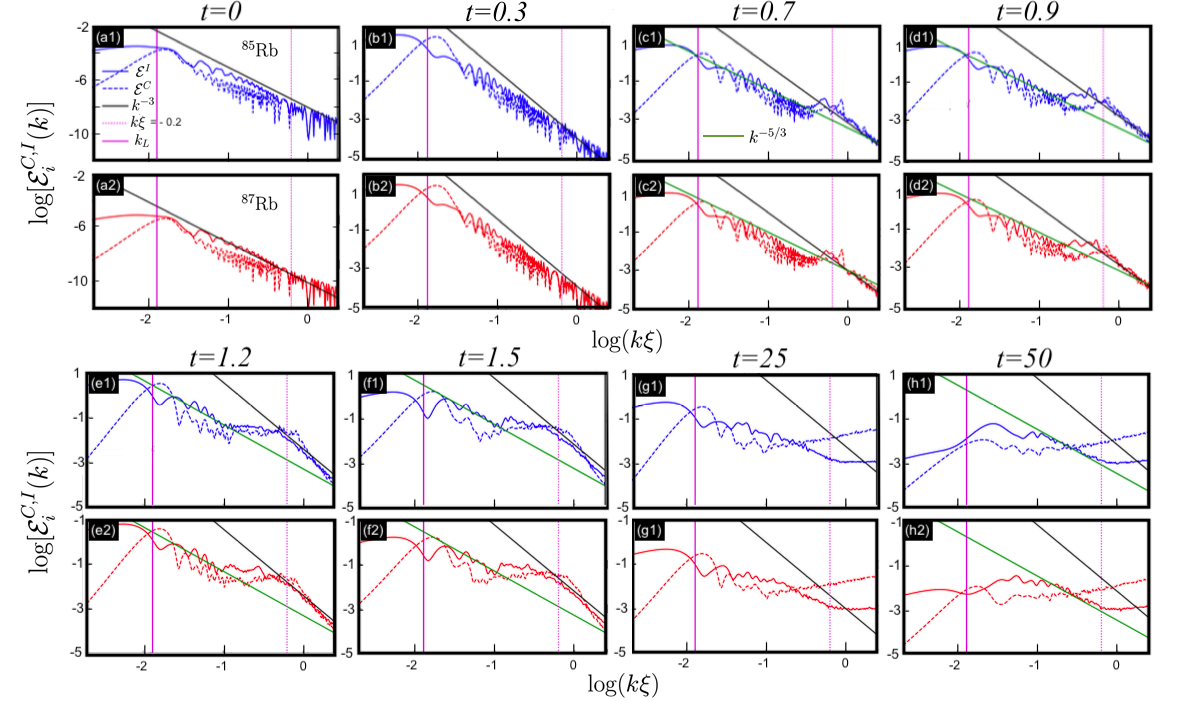}
\caption{Corresponding to the profiles shown in Fig.~\ref{fig5},
the incompressible and compressible kinetic energy spectra of $^{85}$Rb (species 1)
and $^{8t}$Rb (species 2) are presented in log-scale as functions of $\log(k\xi)$ 
(where $\xi$ is the healing length), for given time instants
(indicated at the top of the panels). With the line conventions for all panels 
given in (a1), the expected Kolmogorov behavior stated in (c1) is represented
by the green lines, being more evident for $0.7\le t \le 1.2$.
The vertical solid line refers to the infrared box limit ($k_L\approx 0.13$),
with the vertical dotted line indicating the start of the ultra-violet
($k\approx 2$)
branch region (obtained up to $k_{max}\xi\approx\pi/2$).
The length, energy and time units are, respectively, 
$l_\perp$, $\hbar\omega_\perp$ and $1/\omega_\perp$.
}
\label{fig8}
\end{center}
\end{figure*}
In line with Ref.~\cite{2025Kumar}, we have repeated the analysis for the axial-shaped 
initial configuration, considering different IMQT for the coupled system with 
($^{85}$Rb) and ($^{87}$Rb). The system was initially prepared with $a_{12}=110a_0$, 
larger than $a_{11}=a_{22}=90a_0$. The ground state of the separated mixture is 
evolved in real-time, with the interspecies scattering length interaction quenched to 
$a_{12}=60a_0$. The evolution of the coupled density profiles can be observed in 
Fig.~\ref{fig5}, starting in the homogeneous initial configuration with 
$a_{12}=110a_0$ and $a_{ii}=60a_0$, when $t=0$, followed by a sudden transition to 
a miscible configuration, having $a_{12}=60a_0$ and $a_{ii}=90a_0$, which are kept 
along the evolution.The dynamics of the evolving mixture yield verified interference 
fringes and patterns, which are localized primarily near the trap's boundaries. 
This spatial localization distinguishes it from the behavior observed in Fig.~\ref{fig1}. 

From the above, the initial IMQT is given by $\Delta\delta = 5/9$, much larger than 
the IMQT $\Delta\delta = 0.3$ considered in Ref.~\cite{2025Kumar} for the same kind
of axially-symmetric initial configuration. The corresponding time evolution of the 
density overlap parameter $\Lambda$ is presented in Fig.~\ref{fig6}. From the analysis 
of its asymptotic limit, one can observe that the initial IMQT $\Delta\delta = 5/9$ 
is reflected in an asymptotic oscillating frequency of $\Lambda$ given by 
$\nu_\Lambda\approx 0.5$  (15 cycles in $\Delta t=30$).
This result should be compared with the one obtained in~\cite{2025Kumar}, where for 
$\Delta\delta=0.30$ it was obtained $\nu_\Lambda\approx 0.2$ (6 cycles in $\Delta t=30$). 
From these results for the initial axially symmetric case (two immiscible domains), 
similarly as we have obtained for the case in which the species are in three immiscible domains,
we can establish an approximately linear relationship between the asymptotic frequency 
$\nu_{\Lambda}$ and the initial IMQT, given by $\Delta\delta \approx 1.17 \nu_\Lambda + 0.15$.
Similarly, as we have interpreted before, the factor 1.17 refers to the initial spatial 
configuration in the immiscible regime, with the constant factor 0.15 corresponding to
the threshold IMQT, which emerges in the zero-frequency limit. So, from \cite{2025Kumar}, 
where $\Delta\delta=0.3$, we have $\nu_\Lambda \approx 0.20$; 
whereas, in the present case, for $\Delta\delta=5/9$, we obtain 
$\nu_\Lambda \approx 0.5$.

Apart from the oscillatory behavior shown in Fig.~\ref{fig6}, and consistent with the initial 
IMQT configuration, the vorticity remains approximately stable in the asymptotic regime, 
regardless of the initial immiscible parameterization. The number of vortices, $N_v$, in this 
asymptotic limit can be approximated from the corresponding behavior of the incompressible 
kinetic energy, as previously demonstrated in \cite{2025Kumar} (where $N_v$ was explicitly 
given). Therefore, for the present 
analysis, we can already associate the vorticity in the asymptotic limit with the behavior 
of the incompressible part of the kinetic verified in Fig.\ref{fig7}, where $^{85}$Rb 
is given in panel (a), whereas $^{87}$Rb is in panel (b). These panels are also showing 
for both species, in the asymptotic limit, that the compressible part of the energies, 
related to sound-wave production, are significantly larger (about a factor 5) than 
the incompressible parts.

For the time evolution of the incompressible and compressible parts of the kinetic energy, 
the turbulent flow in the binary mixture can be verified in the initial short-time interval 
of the onset of the instabilities, following approximately the classical Kolmogorov scaling 
$k^{-5/3}$ in the initial short-time interval of the onset of the instabilities. In the 
present case, this scaling is noticed in the time interval $t\approx 0.7$ to 1.2 [See
panels (c$_i$) to (e$_i$) of Fig.~\ref{fig8}], where the $k^{-5/3}$ behavior is indicated 
by the straight-green line.

\section{Conclusion}\label{sec4}
In our present study related to the dynamics and corresponding turbulent behavior of 
the $^{85}$Rb-$^{87}$Rb binary Bose-Einstein condensate mixture in a uniform circular 
box, we reinforce the conclusions previously presented in \cite{2025Kumar} for the 
general qualitative behavior of the dynamics when going from an immiscible to a 
miscible configuration. 
As a supplement to a previous more general investigation, the main outcome we are presenting 
in this contribution refers to the overall time reduction in the dynamics due to the 
larger immiscible to miscible quench transitions, for the two kinds of initial spatial configurations 
considered in Ref.~\cite{2025Kumar}. 
Here, with intraspecies interactions kept constant at $a_{ii}=90a_0$, the interspecies is 
quenched from 100$a_0$ to 60$a_0$ when three spatial domains are considered for the initially 
immiscible configuration; and reduced from 110$a_0$ to 60$a_0$ when it is initially given by 
two spatial axially symmetric configuration.  
As observed, in all the results that we have presented for the time evolution of the
coupled densities, in the immiscible to miscible transition, the dynamics can be 
significantly reduced, emerging in a faster way when considering larger IMQT in 
the initial configuration. Even though they are qualitatively similar, the incompressible 
and compressible parts of the kinetic energy show up specific characteristic behaviors 
previously discussed, with production of vorticity and sound waves, in a faster way 
when considering larger IMQT.
Based on the time evolution of the miscibility parameter $\Lambda$ as the system 
transitions from immiscible to miscible configurations, we can also establish a linear relationship 
between the asymptotic oscillation frequency $\nu_\Lambda$ and the initially imposed 
IMQT $\Delta\delta$. This relationship is given by
$\Delta\delta \approx \alpha \;\nu_\Lambda + 0.15$, 
where $\alpha$ is a parameter that depends on the initial spatial configuration. 
Specifically, in the models that we have considered, we find $\alpha = 1.17$ when 
the immiscible system is localized in two spatial domains, and $\alpha = 0.37$ 
when it is localized in three spatial domains.
Since the constant term 0.15 is common to both cases, we interpret it as corresponding to 
the zero-frequency limit attainable for any IMQT, which appears in the asymptotic regime 
independently of the initial spatial arrangement.

As a final remark, we should note that the current IMQT study of a binary condensed 
system, initiated in Ref.~\cite{2025Kumar}, is still preliminary. Further analytical 
investigation is required, 
particularly through the exploration of different initial conditions for the quenching 
transition of the immiscible mixture. In addition, the asymptotic behavior of the 
compressible kinetic energy — related 
to sound-wave propagation — should be examined in greater detail.

\bmhead{Acknowledgments}
We are grateful for the partial support received from:\\
Funda\c{c}\~{a}o de Amparo \`{a} Pesquisa do Estado de S\~{a}o Paulo (FAPESP)
[Grants 2024/04174-8 (SS and LT) and 2024/01533-7 (AG and LT)], and 
Conselho Nacional de Desenvolvimento Cient\'\i fico e Tecnol\'ogico (CNPq) 
[Grants 303263/2025-3 (LT) and 306219/2022-0 (AG)].
RKK also acknowledges previous partial support from CNPq (Grant 153522/2018-6), 
which facilitated part of this work.

\section*{Declarations}

\begin{itemize}
\item Author contribution: L.T. wrote the main text, which was discussed and 
revised by A.G. and S.S. The numerical data were obtained by R.K.K. and S.S.
\item Data availability: Related material are available from the authors under request.
\item Competing interests: The authors declare that they have no competing interests.
\end{itemize}

\backmatter


\begin{thebibliography}{99}
\bibitem{2025Kumar} R. K. Kumar,  S. Sabari, A. Gammal, and L. Tomio, 
{Rayleigh-Taylor, Kelvin-Helmholtz, and immiscible-to-miscible 
quenching instabilities in binary Bose-Einstein condensates}, 
Phys. Rev. A {\bf 112}, 033312 (2025).

\bibitem{Kolmogorov1941}
A. N. Kolmogorov, 
The local structure of turbulence in incompressible viscous fluid for very 
large Reynolds' numbers, Dokl. Akad. Nauk SSSR {\bf 30}, 301 (1941).

\bibitem{1995Frisch}
U. Frisch, Turbulence: The Legacy of A. N. Kolmogorov
(Cambridge University Press, Cambridge, 1995).

\bibitem{1986Donnelly} R. J. Donnelly and C. E. Swanson, {Quantum turbulence}, 
J. Fluid Mech. {\bf 173}, 387 (1986).

\bibitem{2024Bayocboc} F. A. Bayocboc Jr., J. Dziarmaga, and W. H. Zurek,
Biased dynamics of the miscible-immiscible quantum phase transition in a 
binary Bose-Einstein condensate, Phys. Rev. B {\bf 109}, 064501 (2024).

\bibitem{Mukherjee2020} K. Mukherjee, S. I. Mistakidis, P. G. Kevrekidis, 
and P. Schmelcher, Quench induced vortex-bright-soliton formation in binary 
Bose-Einstein condensates, J. Phys. B: At. Mol. Opt. Phys. {\bf 53}, 055302 
(2020).

\bibitem{2016Eto} 
Y. Eto, M. Takahashi, M. Kunimi, H. Saito, and T. Hirano,
Nonequilibrium dynamics induced by miscible–immiscible transition in 
binary Bose–Einstein condensates, New J. Phys. {\bf 18}, 073029 (2016).

\bibitem{2004Kasamatsu} K. Kasamatsu and M. Tsubota,
Multiple Domain Formation Induced by Modulation Instability in Two-Component 
Bose-Einstein Condensates, Phys. Rev. Lett. {\bf 93}, 100402 (2004).

\bibitem{2019Thiruvalluvar}R.T. Thiruvalluvar, E. Wamba, S. Sabari, 
K. Porsezian, {Impact of higher-order nonlinearity on modulational 
instability in two-component Bose-Einstein condensates}, 
Phys. Rev. E {\bf 99}, 032202 (2019).

\bibitem{2022Bradley} A. S. Bradley, R. K. Kumar, S. Pal, and X. Yu,  
{Spectral analysis for compressible quantum fluids}, 
Phys. Rev. A {\bf 106}, 043322 (2022).

\bibitem{2013Reeves} M. T. Reeves, T. P. Billam, B. P. Anderson,
and A. S. Bradley, {Inverse energy cascade in forced two-dimensional 
quantum turbulence}, Phys. Rev. Lett. {\bf 110}, 104501 (2013).

\bibitem{Bagnato2009}
E. A. L. Henn, J. A. Seman, G. Roati, K. M. F. Magalh\~aes, and V. S. Bagnato, 
 Emergence of turbulence in an oscillating Bose-Einstein condensate,
 Phys. Rev. Lett. {\bf 103}, 045301 (2009). 
 
 \bibitem{2016Tsatsos} M. C. Tsatsos, P. E. S. Tavares, A. Cidrim, A. R. Fritsch, 
M. A. Caracanhas, F. E. A. dos Santos, C. F. Barenghi, and V. S. Bagnato, 
{Quantum turbulence in trapped atomic Bose–Einstein condensates}, Phys. Rep.
{\bf 622}, 1 (2016).

\bibitem{2025Moreno} M. A. Moreno-Armijos, A. R. Fritsch, A. D. García-Orozco, 
S. Sab, G. Telles, Y. Zhu, L. Madeira, S. Nazarenko, V. I. Yukalov, and V. S. Bagnato, 
Observation of Relaxation Stages in a Nonequilibrium Closed Quantum System:
Decaying Turbulence in a Trapped Superfluid, Phys. Rev. Lett. {\bf 134}, 023401 (2025).
\bibitem{2024Madeira} L. Madeira, A. D. García-Orozco, M. A. Moreno-Armijos, 
A. R. Fritsch, and V. S. Bagnato, Universal scaling in far-from-equilibrium quantum systems: 
An equivalent differential approach, Proc. Natl. Acad. Sci. U.S.A. 121, 
e2404828121 (2024).

\bibitem{2023Silva} A. N. da Silva, R. K. Kumar,  A. S. Bradley, 
and L. Tomio, {Vortex generation in stirred binary Bose-Einstein condensates}, 
Phys. Rev. A {\bf 107}, 033314 (2023).

\bibitem{Lauro2024}L. Tomio, A. N. da Silva, S. Sabari, R. K. Kumar,  
{Dynamical vortex production and quantum turbulence in perturbed Bose-Einstein
condensates}, Few-Body Systems {\bf 65}, 13 (2024).

\bibitem{Sabari2024}S. Sabari, R. K. Kumar, L. Tomio, { Vortex dynamics and turbulence in 
dipolar Bose-Einstein condensates}, 
Phys. Rev. A {\bf 109}, 023313 (2024).

\bibitem{2019Kumar} R. K. Kumar,  L. Tomio, and A. Gammal, 
Vortex patterns in rotating dipolar Bose–Einstein condensate mixtures 
with squared optical lattices, 
J. Phys. B:  At. Mol. Opt. Phys. {\bf 52}, 025302 (2019). 

\bibitem{1999Timmermans} E. Timmermans, P. Tommasini, M. Hussein, and A. Kerman, 
{Feshbach resonances in atomic Bose-Einstein condensates}, Phys. Rep. {\bf 315}, 199
(1999).

\bibitem{2017kumar} R. K. Kumar,  P. Muruganandam, L. Tomio, and A. Gammal, 
Miscibility in coupled dipolar and non-dipolar Bose-Einstein condensates,
J. Phys. Commun. {\bf 1}, 035012 (2017). 

\bibitem{2007Lvov} 
V. S. L'vov, S. V. Nazarenko, and O. Rudenko,
Bottleneck crossover between classical and quantum superfluid turbulence,
Phys. Rev. B {\bf 76}, 024520 (2007).

\end{thebibliography}
\end{document}